\documentclass[slac_one]{revtex4}
\usepackage{graphicx}
\usepackage{subfigure}
\usepackage{fancyhdr}
\begin{document}

\title{Prospects for non-standard SUSY searches at LHC} %% Paper title goes here


\author{M. Chiorboli {\em on behalf of the ATLAS and CMS Collaborations}}
\affiliation{Universit\`a di Catania and INFN Sezione di Catania}
\begin{abstract}
New studies of the ATLAS and CMS collaborations are presented on the
sensitivity to searches for non-standard signatures of particular SUSY
scenarios. These signatures include non-pointing photons, as well as pairs
of prompt photons, as expected in GMSB SUSY models, as well as heavy stable
charged particles produced in split supersymmetry models, long lived staus
from GMSB SUSY and long lived stops in other SUSY scenarios. A detailed
detector simulation is used for the study, and all relevant Standard Model
backgrounds and detector effects that can mimic these special signatures
are included. It is shown that already with less than 100 pb$^{-1}$ the
ATLAS and CMS sensitivity will probe an interesting as yet by data
unexplored parameter range of these models.
\end{abstract}

\maketitle

\thispagestyle{fancy}


\section{INTRODUCTION} % Section title should be in all capitals.
In the past few years several studies have been performed by the ATLAS and CMS collaboration to test the capability of their
detectors to observe New Physics with non-standard signatures
(for searches with leptons, jets and missing transverse energy in the final state see S. Caron in these proceedings).
This report focuses on signals of New Physics at the Large Hadron Collider (LHC) with pointing and non-pointing photons and with
Heavy Stable Charged Particles (HSCP).

\section{FINAL STATES WITH POINTING AND NON-POINTING PHOTONS}

Gauge Mediated Supersymmetry Breaking (GMSB)~\cite{GMSB} is a spontaneously broken local supersymmetry. 
%A messenger sector is postulated
%and the breaking is transmitted to the messenger sector to the hidden sector via new interactions, and then trasmitted
%to the visible sector via gauge interactions.
If GMSB events will be produced at the LHC, they will have production mechanisms similar to the ones foreseen for
the Minimal Supergravity (mSUGRA), with large cross sections due to the strongly interacting squarks and gluinos,
and long decay chains, leading to final states with many leptons and jets. The main phenomenological difference with the mSUGRA
is related to the Lightest Supersymmetric Particle (LSP): in GMSB the LSP is the Gravitino, having a very low mass (less than a few KeV's),
while the neutralino or the stau is the Next-to-Lightest Supersymmetric Particle (NLSP). If the neutralino is a NLSP, it can decay to a Gravitino
and a photon; if the stau is a NLSP, it can decay to a Gravitino and a tau. The decay time of such decays can be very long:
depending on the model parameters, it is hence possible to have photons emitted far from the interaction point or stau partially or totally crossing the detector.
The latter scenario will be covered in the section concerning Heavy Stable Charged Particles.

%\begin{figure}[h]
%\includegraphics{figure/ChiDecay.eps}%
%\caption{Feynman Diagram of the decay of a NLSP Neutralino in GMSB.}
%\label{fig:ChiDecay}
%\end{figure}

\subsection{Pointing photons}

\begin{figure}
\centering
%\subfigure[Bild a.] % caption for subfigure a
{
%    \label{fig:AtlasPhotons}
    \includegraphics[width=0.4\textwidth]{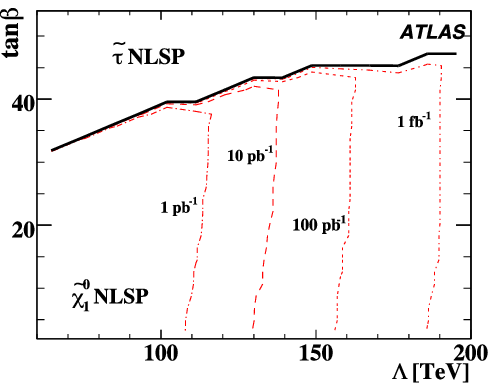}
}
\hspace{1cm}
%\subfigure[Bild b.] % caption for subfigure b
`{
    \includegraphics[width=0.4\textwidth]{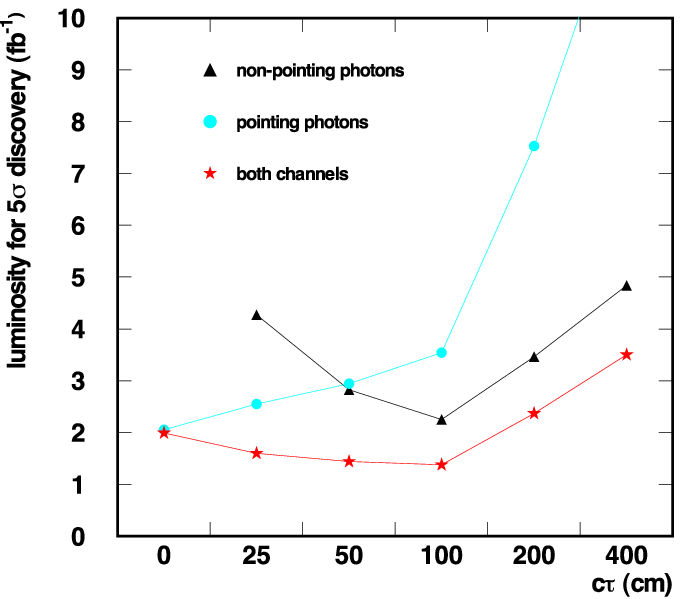}
%    \label{fig:AtlasScan}
}
\caption{Left: 5$\sigma$ discovery regions in the $\Lambda$-$\tan\beta$ plane for different integrated luminosities in an ATLAS analysis.
  Right: Integrated luminosity necessary for 5$\sigma$ discovery in a CMS GMSB analysis. }
\label{fig:Photons} % caption for the whole figure
\end{figure}

If the lifetime of the neutralino is short, the photon arising from its decay is seen as emitted from the interaction point. Considering also the
rest of the event, the final state will be characterized by leptons, jets, missing transverse energy (given by the Gravitino escaping the detector) and
one or two high momentum photons.
In an ATLAS analysis performed with full simulation~\cite{ATLAS_GMSB}, a set of cuts usually adopted for standard SUSY analyses is applied to simulated data
in a preselection stage. The cuts are: $E_T^{miss}>$100~GeV, $E_T^{miss}>$0.2$M_{eff}$, N$_{jets}>$4, $p_{T}^{jets}>$50~GeV/c, $p_{T}^{jet1}>$100~GeV/c.
The analysis is performed at a benchmark point given by the GMSB parameters M=500~TeV/c$^2$, N=1, sgn($\mu$)=+, $\Lambda$=80~TeV, $\tan\beta$=5 and C$_{grav}$=1.
With these parameters the branching ratio of the decay of the neutralino to a photon and a Gravitino is $\sim$97\%, and the photon is expected to originate
close to the interaction vertex ({\em prompt photon}).
%The distribution of number of photons expected after the preselection cuts is shown for signal and background events in Figure~\ref{fig:AtlasPhotons} for
%1~fb$^{-1}$ of integrated luminosity. 
The additional requirement of two photons with p$_T>$20~GeV/c and $|\eta|<$2.5 selects 252.9 signal events and 0.1 background events
for 1~fb$^{-1}$ of integrated luminosity. In this scenario an excess of GMSB events over the backgrounds can be therefore easily observed  
with a much lower integrated luminosity.
To estimate the detector capability to observe signal excess over the expected Standard Model backgrounds, a scan of the GMSB parameter space has been
performed using the Fast Simulation package of the ATLAS collaboration. The result is shown in Figure~~\ref{fig:Photons}. A discovery with even
less than 10~pb$^{-1}$ of integrated luminosity is possible in a favourable region of the parameter space.

%\begin{figure}
%\centering
%\subfigure[Bild a.] % caption for subfigure a
%  {
%    \label{fig:AtlasPhotons}
%    \includegraphics[width=0.45\textwidth]{figure/Nphoton_ink.eps}
%   }
%\hspace{1cm}
%\subfigure[Bild b.] % caption for subfigure b
%`{
%    \label{fig:AtlasScan}
%    \includegraphics[width=0.45\textwidth]{figure/AtlasScan.eps}
%
%\caption{Left:Number of photons after the ATLAS preselection for 1 fb¡Ý1 for signal and background. 
%  Right: 5$\sigma$ discovery regions in the $\Lambda$-$\tan\beta$ plane for different integrated luminosities.}
%\label{fig:AtlasPhotons} % caption for the whole figure
%\end{figure}

\subsection{Non-pointing photons}
If the decay time of the neutralino is long, the photon is emitted far from the interaction point. It is possible to
distinguish this situation from the one described in the previous section by looking at the symmetry of the energy deposit in the
Electromagnetic Calorimeter: an asymmetric ellipse is expected for photons not coming from the interaction vertex.
In a CMS analysis performed with a full simulation of the detector~\cite{CMS_nonPointing} both cases are taken into account
simultaneously. A high momentum photon is required at trigger level, while the offline selection asks for an 
$E_T^{miss}>$160~GeV, at least four jets with $p_{T}^{jets}>$50~GeV/c and $|\eta^{jet1}|<$1.7, and one isolated photon
with $p_{T}^{\gamma}>$80~GeV/c. Two estimators are defined to quantify the asymmetry of the energy deposit of the photon in
the Electromagnetic Calorimeter, and two different bins of these estimators are considered, one optimized for the pointing photons
and one for the non-pointing photons. The integrated luminosity needed to reach a 5$\sigma$ discovery versus the decay time of the
neutralino is shown in Figure~\ref{fig:Photons} for both cases and for the combined analysis. An integrated luminosity
of 1~fb$^{-1}$ is sufficient for the discovery of a neutralino with a c$\tau\sim$~100~cm, in a GMSB scenario with $\Lambda$=180~GeV from
the SPS8 line~\cite{SPS}.

\section{HEAVY STABLE CHARGED PARTICLES}

As said in the previous sections, some parameter regions of GMSB foresee the possibility of a long-lived stau crossing partially of completely
the detector. Other models Beyond the Standard Model predict the production of Heavy Stable Charged Particles at LHC: lepton-like HSCP can be produced
as Kaluza-Klein tau resonances in Universal Extra Dimension (UED)~\cite{UED},
while heavy stable hadrons (called R-Hadrons) can be produced by heavy stable stops in some MSSM scenarios~\cite{stops} 
or by heavy stable gluinos in Split-SUSY~\cite{SplitSusy}. 
In all these models the HSCP have masses of the order of hundreds of GeV, but have low velocities, with $\beta$ values being significantly lower than one. 
In a recent CMS analysis~\cite{CMS_HSCP} all the models above are considered, but the study is performed in a model independent way, just looking at the
heavy particle crossing the detector, without any assumption on the rest of the event.

\subsection{Detection of HSCP}

Since HSCP are non relativistic, they are characterized by large ionization and large time of flight. These features can be exploited for their identification.
Lepton-like HSCP simply behave like ``heavy muons'' crossing the detector. R-hadrons are instead modeled as a gluino or a stop bound 
to a quark-antiquark pair (R-mesons) or
to three quarks (R-barions). This modelization has two important consequences:
in the interaction models of the R-Hadrons the heavy particle acts as a spectator bringing most of the kinetic energy, while
the interactions are given by the light quark cloud: no showers are therefore foreseen in the calorimeters. In addition the charge of the R-hadrons can flip in their
travel through the detector material, since one of the light quarks can be exchanged with one of the quarks of the material.
Measuring the momentum and the velocity of the HSCP it is possible to get their mass; the HSCP can be hence detected looking at slow particles having large mass.
The momentum can be extracted from the bending of the tracks reconstructed in the tracker or in the muons spectrometer of the CMS and ATLAS detectors,
while the velocity can be measured from the time of flight in the muon spectrometer of from the dE/dx in the tracker, as described in the following section.
Two methods can be followed to trigger HSCP events: looking at the particle itself using a muon trigger, or looking at the rest of the event, using a jet or missing
transverse energy trigger. The former case is completely model independent, but very slow particles can be assigned to the wrong bunch crossing, leading to
a reduction in the efficiency: the minimum particle $\beta$ to have a muon trigger signal sinchronyzed with the signals produced by the particle in the
inner detectors is about 0.65 and 0.75 for CMS and ATLAS, respectively. 
The jet and missing transverse energy triggers do not suffer from wrong bunch crossing assignment but are less model independent.
Other experimental issues have to be taken into account in an HSCP analysis: the charge flip of a particle can make a muon trajectory unmatched with the tracker one,
and can make neutral born R-hadrons to acquire a charge while crossing the detector, yielding a track reconstructed in the muon spectrometer but not
in the tracker.

\subsection{HSCP analysis}

\begin{figure}
\centering
%\subfigure[Bild a.] % caption for subfigure a
{
%    \label{fig:AtlasPhotons}
    \includegraphics[width=0.4\textwidth]{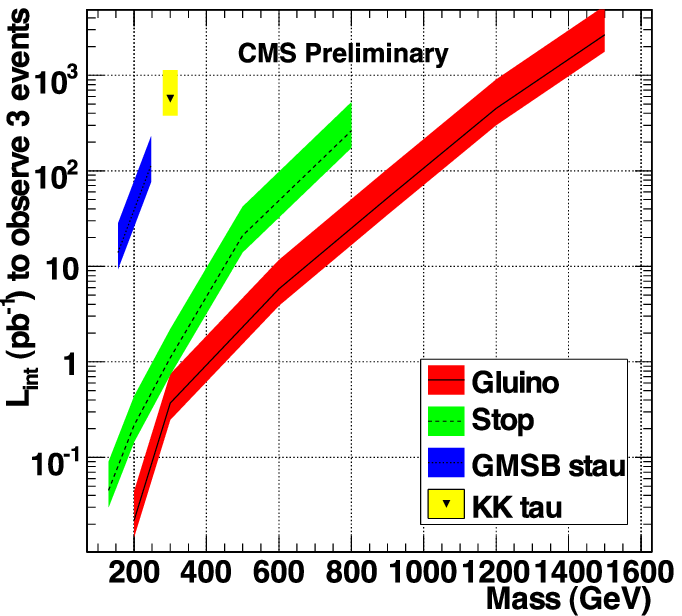}
}
\hspace{1cm}
%\subfigure[Bild b.] % caption for subfigure b
`{
%    \label{fig:AtlasScan}
    \includegraphics[width=0.4\textwidth]{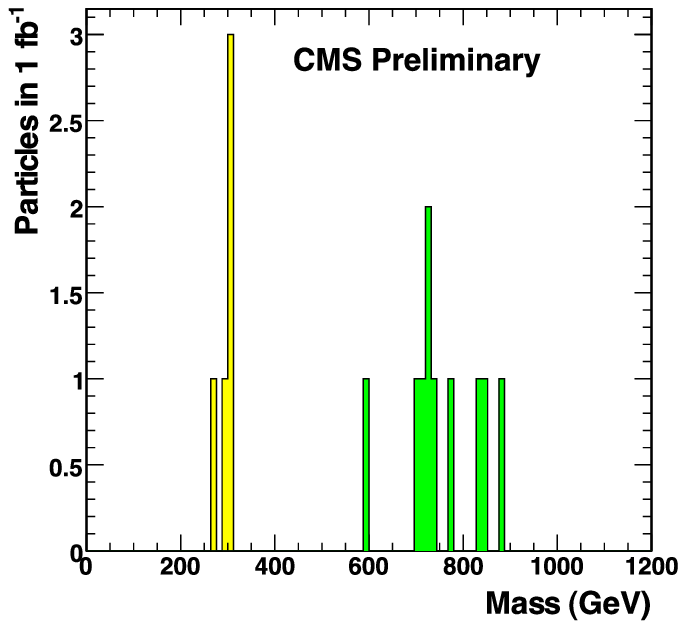}
}
\caption{Left:integrated luminosity needed for 3 signal events for four different signal models, as a function of the HSCP mass. 
  Right: reconstructed mass distribution with 1~fb$^{-1}$ of integrated luminosity for a 300~GeV/c$^2$ KK tau (yellow) and for a 800~GeV/c$^2$ stop (green).}
\label{fig:CMS_HSCP_curves} % caption for the whole figure
\end{figure}

In a CMS analysis performed with a full simulation of the detector, the velocity of the particles are measured from the time of flight in the muon chambers
and from the dE/dx in the tracker. A dedicated GEANT4 package has been developed to simulate the interaction of HSCP's with matter~\cite{CustomPhysics},
and new reconstruction techniques have been implemented in the CMS reconstruction software for the estimate of their velocities. The
considered Standard Model backgrounds are W+jets, Z+jets, $t\overline{t}$+jets as well as muons from Drell-Yan, Z, W, bottomonia, charmonia and jets. 
The time of flight is measured exploiting the excellent ($\sim$~1~ns) time resolution of the Drift Tubes (DT) mounted in the CMS muon spectrometer, which allows
the distinction between relativistic and non relativistic particles. The DT are normally used to estimate the position of tracks via measurements of the drift time
of ionization. Non relativistic particle velocities can be extracted from a fit if the drift time is left as a free parameter when the information
from the staggered drift tubes is combined to get the crossing point of the particle. Main backgrounds are given by cosmic rays and by tails in the measured velocity distribution of true muons; with real data the latter ones can be estimated using Z$\rightarrow\mu\mu$ decays.
The cosmic background can be strongly suppressed combining the tracks
reconstructed in the DT with the ones reconstructed in the tracker.
The velocity is also obtained from the dE/dx of the particle in the tracker inverting the common Bethe-Block formula. Low momentum protons and kaons in real data
can be used to obtain the coefficient of proportionality between the velocity and the dE/dx, while minimum ionizing particles (MIP's) can be used to estimate
the absolute dE/dx scale and the tails. Main background for this measurement is given by tails in dE/dx for Standard Model particles. 
In the CMS analysis described here the requirement that both the velocities measured with the two methods are less than 0.8c, in addition to the requirement that
the average of the two measured masses is greater than 100~GeV/c$^2$, select zero background events.
In Figure~\ref{fig:CMS_HSCP_curves} the discovery curves refer to the luminosity needed to collect 3 signal events for the various models considered.
The error bars correspond to a systematic uncertainty of about 50\%, which in this study is dominated by the muon trigger efficiency. 
An integrated luminosity lower than 100~pb$^{-1}$ is sufficient up to an HSCP mass of about 1~TeV/c$^2$. In Figure~\ref{fig:CMS_HSCP_curves}
two different HSCP reconstructed mass peaks are shown for 1~fb$^{-1}$ of integrated luminosity.

\section{CONCLUSIONS}
The ATLAS and CMS collaborations have been developing new techniques to observe new physics through non standard final states, like pointing and non-pointing
photons or heavy stable charged particles. The simulation studies are promising, showing the capability of both the detectors to discover new physics,
in favourable scenarios, with less than 100~pb$^{-1}$ of integrated luminosity.

\begin{acknowledgments}
The author wish to thank A. De Roeck and G. Polesello for providing useful material, A. Giammanco and M. Galanti for help with technical questions, S. Costa for reading the manuscript.
\end{acknowledgments}

\end{document}